\documentclass[aps,prl,twocolumn,showpacs]{revtex4-1}  % for review and submission

\usepackage{dcolumn}% Align table columns on decimal point
\usepackage{bm}% bold math
\usepackage{natbib}
\usepackage{graphicx}
\usepackage{pdfpages}
\usepackage{eso-pic}
\usepackage{everyshi}
\usepackage{multido}

\begin{document}

%\preprint{APS/123-QED}

\author{Thomas Hofmann}

\affiliation{Institute of Experimental and Applied Physics, University of Regensburg, D-93053 Regensburg, Germany}

\author{Florian Pielmeier}

\affiliation{Institute of Experimental and Applied Physics, University of Regensburg, D-93053 Regensburg, Germany}

\author{Franz J. Giessibl}
\email{franz.giessibl@ur.de}

\affiliation{Institute of Experimental and Applied Physics, University of Regensburg, D-93053 Regensburg, Germany}

\date{\today}% It is always \today, today,

\title{\textbf{Chemical and Crystallographic Characterization of the Tip Apex in Scanning Probe Microscopy}}

\begin{abstract}
The apex atom of a W scanning probe tip reveals a non-spherical charge distribution as probed by a CO molecule bonded to a Cu(111) surface [Welker \textit{et al.} Science, 336, 444 (2012)]. Three high-symmetry images were observed and related to three low-index crystallographic directions of the W bcc crystal. Open questions remained, such as the detectability of a contamination of W tips by sample material (here Cu), and the applicability of the method to distinguish other atomic species. In this work, we investigate bulk Cu and Fe tips. In both cases, we can associate our data with the fcc (Cu) and bcc (Fe) crystal structures using a simple electrostatic model that is based on the partial filling of d orbitals.
\end{abstract}

\maketitle

The front atom of the tip in a scanning probe microscope is important - it can be compared to the objective lens in an optical microscope. In scanning tunneling microscopy (STM), the tip can often be treated as spherically symmetric (Tersoff-Hamann s-wave model~\cite{Tersoff1983}). In atomic force microscopy (AFM), the front atom strongly affects both the appearance of imaged atoms~\cite{Welker2012} as well as force spectroscopy~\cite{Welker2013}. An indirect method to determine the tip cluster is the comparison of experimental results with \textit{ab initio} calculations for different realistic tip models~\cite{Sugimoto2007,Teobaldi2011,Ternes2011}. Another possibility is the combination of field ion microscopy, which can resolve the atomic structure of a sharp asperity~\cite{Mueller1951}, with an AFM~\cite{Falter2013,Paul2013}. A common method is the functionalization of the tip apex with atoms or molecules picked up from the surface~\cite{Bartels1998,Gross2009b,Pavlicek2011,Boneschanscher2012,Kichin2013,Mohn2013}. Carbon monoxide (CO) is commonly used, because its confined and closed-shell electronic structure makes it an inert probe for high-resolution atomic force imaging~\cite{Gross2009b,Sun2011,Pavlicek2011,Boneschanscher2012}. Recently, a CO molecule, adsorbed on a Cu(111) surface, has been used to probe the apex atom of a W tip~\cite{Welker2012} - a technique introduced as \textbf{CO} \textbf{F}ront atom \textbf{I}dentification (COFI)~\cite{Welker2013}.

In this Letter, we show that COFI images and force spectroscopy can be used to distinguish Cu front atoms from W and Fe front atoms and to determine their angular orientation. We further introduce an electrostatic model that relates the symmetries of experimental COFI images obtained with the Cu tips to the partial filling of d orbitals in the apex atom.

\begin{figure*}[bt]
 \includegraphics[width=172mm]{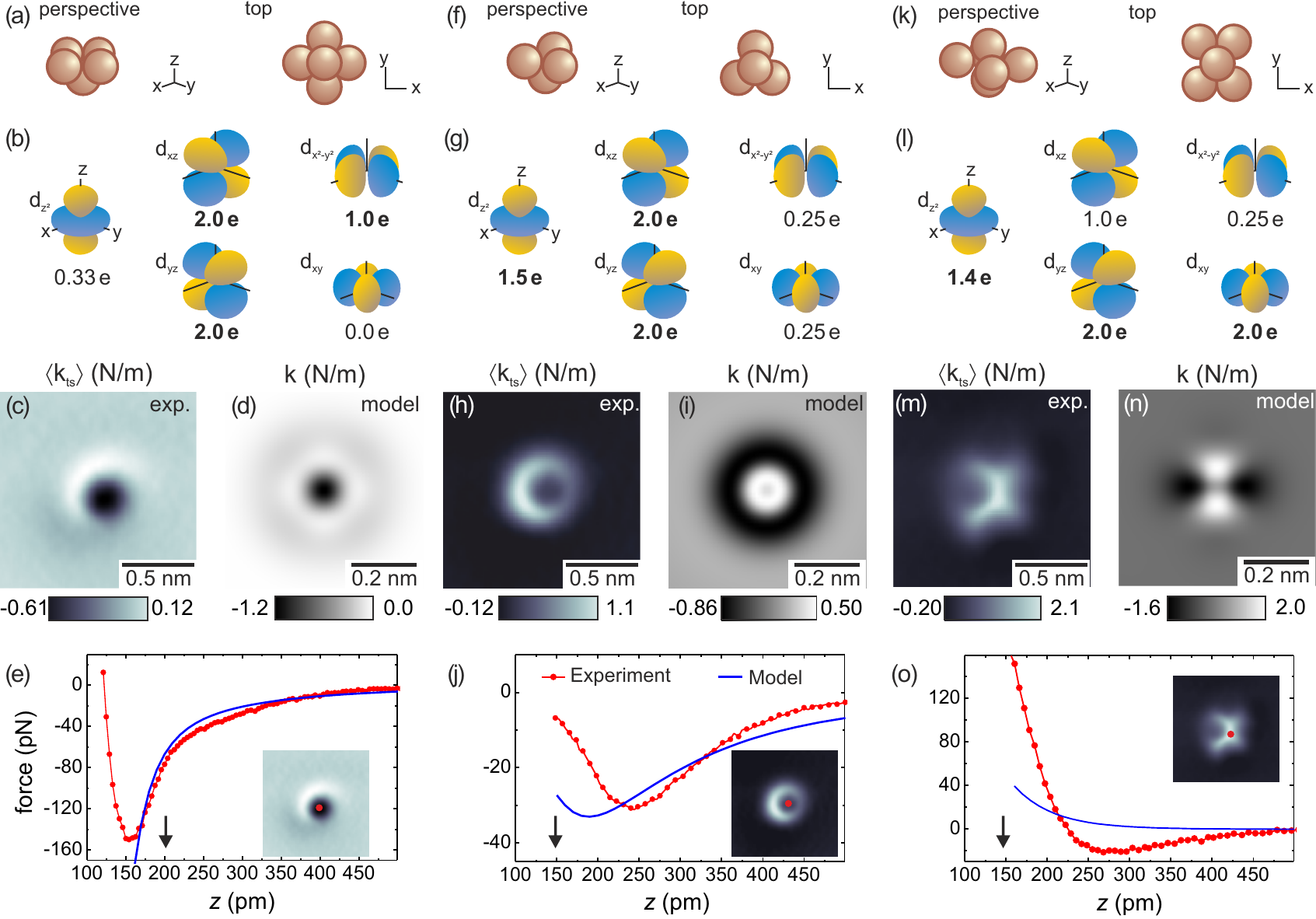}
\caption{(color online). Cu tips yield three high-symmetry COFI images (c, h and m), which can be assigned to Cu tip clusters pointing into the $\langle 100 \rangle$ (a), $\langle 111 \rangle$ (f) and $\langle 110 \rangle$ directions (k). To justify the assignment, we develop a model for the tip atom and calculate its electrostatic interaction with the dipole of the CO molecule. The model tip atom consists of a positive charge in the center, surrounded by unequally occupied 3d orbitals. Their occupation (b, g and l) is determined by their angular overlap with the nearest neighbors. The bold labels refer to the three orbitals with the highest occupation. The central charge is adjusted to optimize the fit between model and experimental results for the force versus distance behavior (e, j and o) and the contrast in the model images, which are calculated at a core-core distance of $200$ (d), $145$ (i) and $160\,\text{pm}$ (n). The black arrows in the force versus distance graphs indicate the distances of the COFI images. The experimental distance is determined from the conductance of the tunneling gap \cite{Supplemental_Info}. In the model, the $z$-value reflects the core-core distance between the Cu tip atom and the oxygen~\cite{Supplemental_Info}.}
\label{fig_Cu_diff_Orient}
\end{figure*}

All measurements were performed in a low temperature STM-AFM operated at $4.4\,\text{K}$ (LT STM/AFM, Omicron Nanotechnology, Taunusstein). The microscope is equipped with a qPlus sensor~\cite{Giessibl1998} and operated in the frequency modulation mode ($A=50\,\text{pm}$). We used qPlus sensors with tips etched from polycrystalline Cu ($99.95\,\%$), Fe ($99.998\,\%$) and W ($99.95\,\%$) wires. The measurements were conducted on a Cu(111) sample, which was covered by about $0.01\,\text{ML}$ CO. Prior to the measurements on the CO/Cu(111) sample, all tips were cleaned \text{in situ} by field evaporation. The apex of the tip was altered by poking it into the Cu surface between image acquisition. For modifying only the front-most part of the tip apex, the tip was approached by a few hundred picometers from the STM setpoint towards the clean Cu surface and then retracted while a bias between $0.5$ and $5\,\text{V}$ was applied~\cite{Welker2012,Supplemental_Info}. We call this procedure a gentle poke, in contrast to a strong poke where the tip is indented several nanometers into the surface. The latter is expected to result in a complete covering of the tip apex with sample material~\cite{Hla2004,Limot2005_2}.

During the measurements with the bulk Cu tips, the tip apex was modified repeatedly by strong and gentle pokes. After each poke, the tip atom was characterized by COFI~\cite{Welker2012,Welker2013}. This involves the tip being scanned at close distance over the CO molecule, while the averaged force gradient $\langle k_{\text{ts}} \rangle$ is recorded. These close $\langle k_{\text{ts}} \rangle$~maps are referred to as \lq COFI images\rq{} in the following. Among these COFI images~\cite{Supplemental_Info} three high-symmetry images are identified (Fig.~\ref{fig_Cu_diff_Orient}). Two are circular symmetric [Fig.~\ref{fig_Cu_diff_Orient}(c) and \ref{fig_Cu_diff_Orient}(h)], the third one shows a twofold symmetry [Fig.~\ref{fig_Cu_diff_Orient}(m)]. From the different contrast in the circular symmetric images, it is already apparent that they represent two different tip clusters. In addition, we obtained force versus distance curves at certain positions in the COFI images. Those recorded in the center of the circular symmetric images [Fig.~\ref{fig_Cu_diff_Orient}(e) and \ref{fig_Cu_diff_Orient}(j)] exhibit a large difference in the magnitude of the force ($150$ vs. $30\,\text{pN}$) and in the $z$ position of the attractive minimum and confirm that the COFI images represent two different tip clusters.

Calculations have shown that bulk tungsten displays a charge density that is significantly larger towards the neighboring atoms~\cite{Posternak1980,Mattheiss1984,Wright2011}, while calculations for
Cu predict a much more uniform electron distribution~\cite{Fong1975,Gay1977,Euceda1983,Euceda1983_2}. The strong angular force variations between a W tip and a CO molecule adsorbed on Cu(111) can be explained by the charge density of bulk W~\cite{Welker2012}, while the experimental angular dependencies between Cu tips and CO molecules presented here cannot be explained by the bulk charge density of Cu. Therefore we develop a model for the Cu tip atom that relies on a partial depletion of the d-shell, leading to non-spherical electron distributions for tip clusters pointing into the high-symmetry orientations of a fcc crystal $\langle 100 \rangle$, $\langle 111 \rangle$ and $\langle 110 \rangle$ (Fig.~\ref{fig_Cu_diff_Orient}). The model explains the experimental observations of toroidal COFI images [Fig.~\ref{fig_Cu_diff_Orient}(c) and \ref{fig_Cu_diff_Orient}(h)] for Cu$\langle 100 \rangle$ and Cu$\langle 111 \rangle$ tips and \lq rooftop\rq-type COFI images for Cu$\langle 110 \rangle$ tips [Fig.~\ref{fig_Cu_diff_Orient}(m)].

Density functional calculations of metal tips confirm that the Smoluchowski effect~\cite{Smoluchowski1941} leads to a surface dipole at the tip atom~\cite{Teobaldi2011}. For a Cu adatom on a Cu(111) surface the same effect has been reported~\cite{Limot2005_2} and calculations show that the occupation of the 4s and 3d~orbitals is reduced compared to an isolated Cu atom~\cite{Rodriguez1994_2}. In a homogeneous electric field, the 3d orbitals are degenerate and the field would deplete (or fill) the orbitals equally. However, the field around a Cu tip atom is expected to be strongly inhomogeneous due to the partially unscreened nuclei of the nearest neighbors ~\cite{Rao1971,Herbst1977} and the energy levels of the 3d orbitals split up according to crystal field theory~\cite{VanVleck1932}. This leads to an unequal depletion of the 3d orbitals and correspondingly, a non-spherical charge density around the Cu tip atom.

Following these arguments, we model the electron distribution at the Cu tip atom by 3d orbitals. Contributions of s or p states are not included. The occupation of the 3d orbitals is estimated by the angular overlap of each orbital with all nearest neighbors~\cite{Schaeffer1965}, assuming that orbitals with a large overlap are lower in energy and, hence, carry a higher occupation (for more details and a discussion of the occupation numbers see~\cite{Supplemental_Info}). The occupation of the orbitals determines the basic features found in the COFI images: In the case of the Cu$\langle 100 \rangle$ tips [Fig.~\ref{fig_Cu_diff_Orient}(b)] the electron distribution has the shape of a torus because the $\text{d}_{\text{xz}}$ and $\text{d}_{\text{yz}}$ orbitals are fully occupied. The same holds true for the Cu$\langle 111 \rangle$ tips [Fig.~\ref{fig_Cu_diff_Orient}(g)] but the torus shows a shallower minimum in the center than the Cu$\langle 100 \rangle$ tips because the $d_{\text{z}^2}$ orbital is also highly populated. The asymmetry in the occupation of the $\text{d}_{\text{xz}}$ and $\text{d}_{\text{yz}}$ orbitals found for the Cu$\langle 110 \rangle$ tips [Fig.~\ref{fig_Cu_diff_Orient}(l)] results in a twofold symmetry which is reflected in the COFI image in Fig.~\ref{fig_Cu_diff_Orient}(m).

For a quantitative comparison of the experiment and the tip model, the interaction of the electron distribution and the CO molecule is estimated by electrostatics, as discussed in Ref.~\cite{Welker2012}. The CO molecule is considered as a dipole with a negative charge on the oxygen atom and a realistic dipole moment of about $0.1\,\text{e} \cdot 100\,\text{pm}$~\cite{Zuo2009,Feng2011}. The negative charge of the electron distribution at the tip atom is compensated by a positive charge $Q_{\text{core}}$ in the center. In order to approximately match the contrast in the COFI images and the characteristics of the force versus distance curves, the magnitude of the positive charge is adjusted. As a result, the net charge of the tip atoms $\Delta=Q_{\text{core}}-Q_{\text{3d elec.}}$ is positive for all tip orientations, with $\Delta_{100}=+0.25\,\text{e}$, $\Delta_{111}=+0.52\,\text{e}$ and $\Delta_{110}=+0.04\,\text{e}$. Calculations of the dipole moment of a metal tip atom yield a value of approximately $0.4\,\text{e}\cdot 100\,\text{pm}$~\cite{Teobaldi2011}. This corresponds to a charge of $+0.2\,\text{e}$ at the tip atom assuming a distance of $200\,\text{pm}$ between the charges, indicating that our estimative model returns a realistic net charge at the tip atom.

The calculated image of the Cu$\langle 100 \rangle$ tip shows a toroidal symmetry with a strong attraction in the center [Fig.~\ref{fig_Cu_diff_Orient}(d)], whereas the image of the Cu$\langle 111 \rangle$ tip only reveals a small attractive dip [Fig.~\ref{fig_Cu_diff_Orient}(i)]. As the experimental toroidal COFI images fall into two categories (large dip in Fig.~\ref{fig_Cu_diff_Orient}(c) and small dip in Fig.~\ref{fig_Cu_diff_Orient}(h)), they can be unambiguously identified as Cu$\langle 100 \rangle$ and Cu$\langle 111 \rangle$ tips with the help of the model. The calculation for the Cu$\langle 110 \rangle$ tip [Fig.~\ref{fig_Cu_diff_Orient}(n)] yields a twofold symmetric image similar to the COFI image in Fig.~\ref{fig_Cu_diff_Orient}(m). This result supports the assignment based on the symmetry. Notably, the experimental tip images are more than 2.5 times the size of the calculated images. This is explained by the bending of the CO molecule that is known to magnify the image of the tip atom~\cite{Welker2013} or the length of intermolecular bonds~\cite{Gross2012} and which is not considered in the calculations.

\begin{figure}[bt]
\includegraphics[width=86 mm] {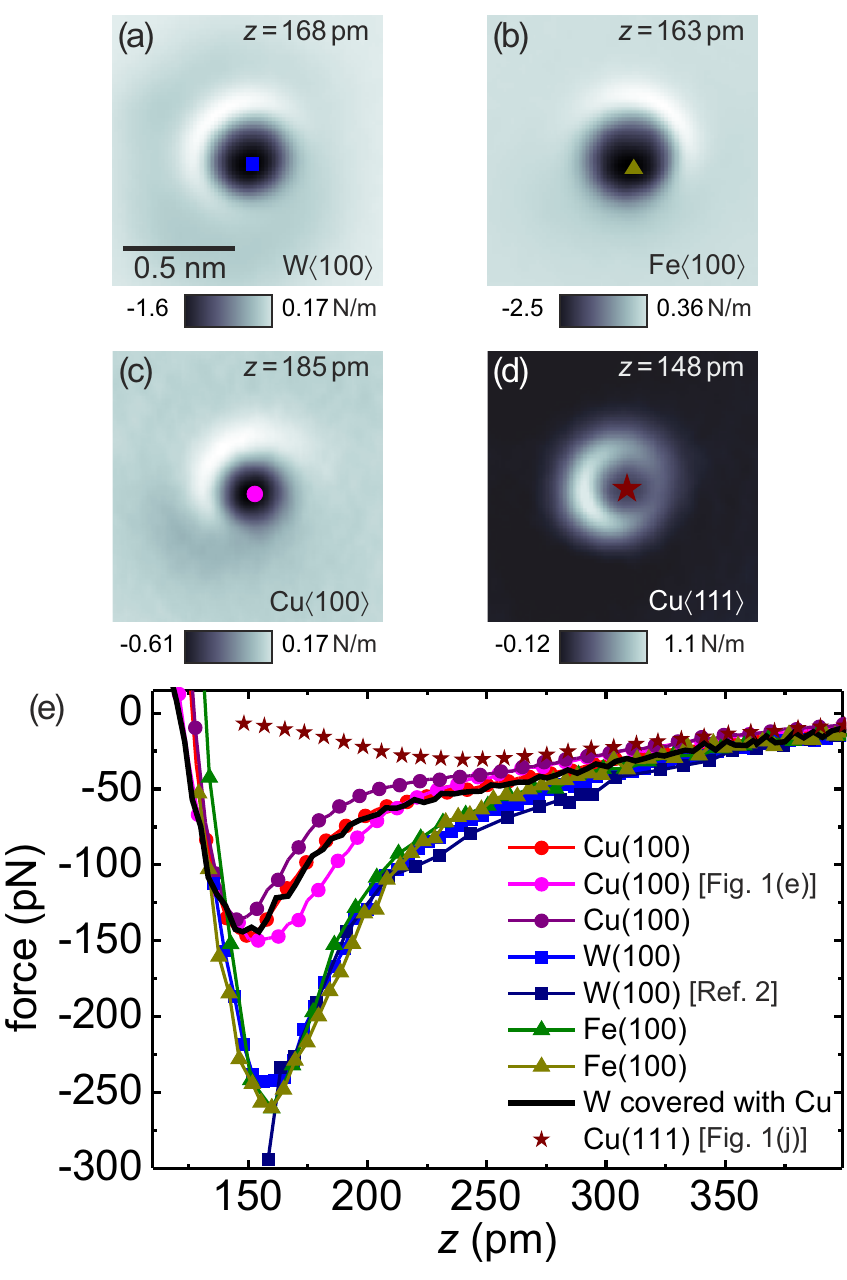}
\caption{(color online). Comparison of W (a), Fe (b) and Cu tips (c and d), which show a circular symmetry in the COFI image [c and d are the same as in Fig.~\ref{fig_Cu_diff_Orient}(c) and~\ref{fig_Cu_diff_Orient}(h)]. e) Force versus distance curves of W$\langle 100 \rangle$ (squares), Fe$\langle 100 \rangle$ tips (triangles), Cu$\langle 100 \rangle$ (circle) and Cu$\langle 111 \rangle$ tips (stars). The solid (black) line represents a W tip, which is contaminated with Cu.}
\label{fig_Tip100}
\end{figure}

\begin{figure}[bt]
\includegraphics[width=86 mm] {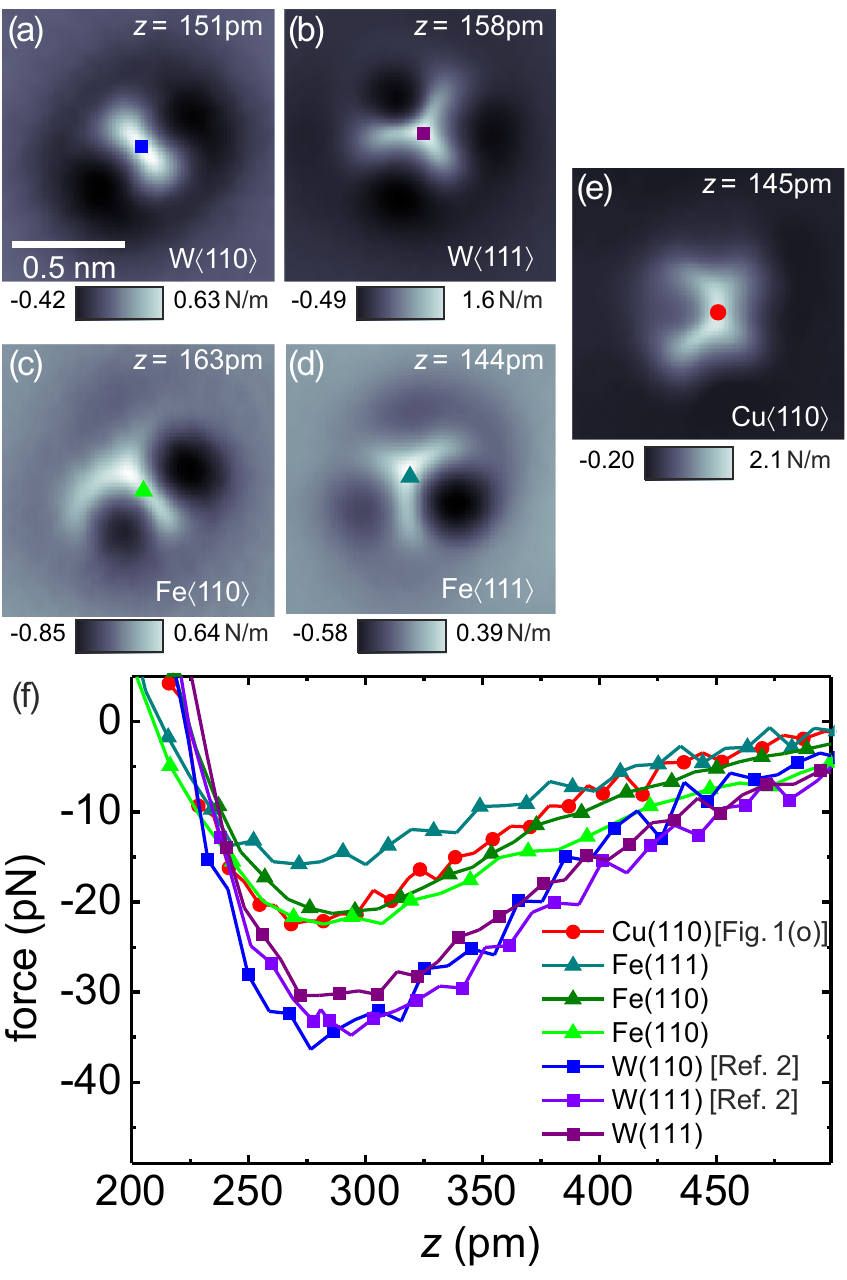}
\caption{(color online). Comparison of W, Fe and Cu tips revealing  two- or threefold symmetry. a-e) COFI images of the corresponding W (a, b), Fe (c, d) and Cu tips (e) (a uses the same raw data as Fig. 1H in~\cite{Welker2012}, e is the same as in Fig.~\ref{fig_Cu_diff_Orient}(m)). f) Force versus distance curves of several W, Fe and Cu tips.}
\label{fig_Tip111_110}
\end{figure}

The measurements with Fe tips again reveal three high-symmetry COFI images [Figs.~\ref{fig_Tip100}(b), \ref{fig_Tip111_110}(c) and~\ref{fig_Tip111_110}(d)]. These exhibit the same symmetries as the W tips [Figs.~\ref{fig_Tip100}(a), \ref{fig_Tip111_110}(a) and~\ref{fig_Tip111_110}(b)] and are therefore assigned to tips oriented in $\langle 100 \rangle$, $\langle 110 \rangle$ and $\langle 111 \rangle$ direction. Similar to W, Fe has a partially filled d shell and a bcc bulk crystal structure, suggesting a higher electron density in the $\langle 111 \rangle$ directions, where the nearest neighbors are located. However, calculations of the total charge density in Fe are not conclusive. In Refs.~\cite{Wang1981,Wu1993} a spherical charge density around the Fe atoms is revealed and no increase in direction of the nearest neighbors is resolved, while a more recent calculation reports a non-spherical electron density~\cite{Jones2008}.

In order to evaluate the potential of the COFI method to distinguish between different chemical species at the tip apex, we compare COFI images and force versus distance curves of Cu, Fe and W tips. The comparison is divided into tips which reveal circular symmetric COFI images (Fig.~\ref{fig_Tip100}), and tips that exhibit a two- or threefold symmetry (Fig.~\ref{fig_Tip111_110}).

Toroidal symmetric COFI images are observed for W, Fe and Cu tips oriented in $\langle 100 \rangle$ direction, and a Cu$\langle 111 \rangle$ tip [Fig.~\ref{fig_Tip100}(a)-(d)]. The Cu$\langle 111 \rangle$ tip can easily be distinguished from all other tips by the contrast in the COFI images and its force versus distance curve [Fig.~\ref{fig_Tip100}(e)]. The COFI images of the  $\langle 100 \rangle$ tips do not allow a discrimination. The force versus distance curves obtained with three different Cu$\langle 100 \rangle$ tips, however, show a significantly smaller maximal attractive force ($130-150\,\text{pN}$, variation explained in~\cite{Supplemental_Info}) than those obtained with Fe$\langle 100 \rangle$ and W$\langle 100 \rangle$ tips ($\approx 250\,\text{pN}$). It is therefore possible to distinguish Cu from Fe and W tips for the $\langle 100 \rangle$ orientations. To support this finding, we modified a W tip apex by several hard pokes into the Cu surface. The COFI image after the modification revealed the image of a tip oriented in $\langle 100 \rangle$ direction. The comparison of its force versus distance curve [black line in Fig.~\ref{fig_Tip100}(e)] with curves of Cu$\langle 100 \rangle$ and W$\langle 100 \rangle$ tips clearly reveals a decoration of the W tip apex with Cu. This is expected for such tip treatment~\cite{Hla2004,Limot2005_2} and proves the capability of COFI to identify a covering of the tip apex with Cu.

In Fig.~\ref{fig_Tip111_110}, examples of COFI images of W, Fe and Cu tips with two- and threefold symmetry are presented. COFI images of W, Fe and Cu tips, oriented in $\langle 110 \rangle$-direction are all twofold symmetric, but the image of the Cu tip [Fig.~\ref{fig_Tip111_110}(e)] is significantly different from those of W [Fig.~\ref{fig_Tip111_110}(a)] and Fe [Fig.~\ref{fig_Tip111_110}(c)]. The Cu tip image features a repulsive part with the shape of a \lq rooftop\rq{} with four shallow minima, whereas the W/Fe tip images exhibit a repulsive bar with two highly attractive minima. Therefore a Cu$\langle 110 \rangle$ tip can be unambiguously identified by the COFI image. As expected, W and Fe tips cannot be distinguished from the qualitative COFI image, as both are bcc crystals.

However, the force versus distance curves of the two- and threefold symmetric tips [Fig.~\ref{fig_Tip111_110}(f)] recorded at the repulsive maximum in the COFI images do allow to distinguish W and Fe tips. We find that curves recorded with W$\langle 110 \rangle$ and W$\langle 111 \rangle$ tips are almost identical, exhibiting a force minimum between $30$ and $35\,\text{pN}$. The curves of the Fe$\langle 110 \rangle$ and Fe$\langle 111 \rangle$ tips, on the other hand, show a smaller attractive minimum of $15$ to $22\,\text{pN}$. Therefore Fe and W can be distinguished with the help of the force versus distance curves. The greater deviation between the force minimum of the Fe$\langle 110 \rangle$ and Fe$\langle 111 \rangle$ tips is probably due to a greater misalignment to the precise $\langle 110 \rangle$ and $\langle 111 \rangle$ orientations. The angular deviations are smaller for the W tips shown in Figs.~\ref{fig_Tip111_110}(a) and \ref{fig_Tip111_110}(b). For details on the effect of angular alignments, see Fig. S\,9 in Ref.~\cite{Welker2012}.

In conclusion, we report three high-symmetry COFI images for Cu and Fe, which can be related to the high-symmetry directions of the fcc and bcc crystals. We therefore suggest that in almost all cases the tip clusters exhibit a bulk-like crystal structure. A tip model based on the partial depletion of selected d orbitals can qualitatively explain the observed COFI images. However, for a quantitative description \textit{ab initio} calculations may be helpful, similar to those calculated for W tips~\cite{Wright2013}. The comparison of the results for W, Fe and Cu tips reveals that COFI is a powerful method to distinguish between the orientation as well as the chemical species of the tip atom. The method has been proven to be useful for measurements on various samples like Si~\cite{Welker2013}, NiO~\cite{Pielmeier2013} and graphene~\cite{Hofmann2013}.

\begin{acknowledgments}

The authors thank A. J. Weymouth, L. Gross and J. Repp for discussions and kindly acknowledge financial support from the Deutsche Forschungsgemeinschaft (Grant No. GRK 1570 and SFB 689).
\end{acknowledgments}

%\bibliography{BibFile_07012014}
\bibliographystyle{apsrev4-1}

\clearpage
%\setboolean{@twoside}{false}
%\includepdf[pages=1-14, fitpaper]{Supplemental_Information.pdf}

 \multido{\i=1+1}{14}{%
       \begin{figure}
       \center
       \vspace{-2cm}\includegraphics[page=\i,scale=0.9]{Supplemental_Information.pdf}
       \end{figure}
        \newpage
        \pagenumbering{gobble}
    }

\end{document}